\documentstyle[12pt]{article}
\def\bm#1{{\mbox{\boldmath $#1$}}}
\begin{document}
\pagenumbering{arabic}
\begin{flushright}
\baselineskip=16pt
{\footnotesize DAMTP--97-133}\\
{\footnotesize FERMILAB--Conf-97/391-A}\\
{\footnotesize hep-th/9711195}
\end{flushright}
\renewcommand{\thefootnote}{\fnsymbol{footnote}}
\footnotesep=14pt
\vspace{0.15in}
\baselineskip=24pt
\begin{center}
{\Large\bf Towards a String Formulation of}\\
{\Large\bf Vortex Dynamics\footnote{\baselineskip=14pt\noindent
Talk presented by O.T.~at the International Workshop on Solitons,
their Properties, Dynamics, Interactions, and Applications in
Kingston, Ontario, Canada, 20-26 July 1997. To appear in the
proceedings.}}\\
\baselineskip=16pt
\vspace{0.75cm}
\addtocounter{footnote}{1}
{\bf Elsebeth Schr\"{o}der\footnote{Electronic address:
{\tt schroder@nbi.dk}}}\\ {
{\small\em Physikalisches Institut der Universit\"at Bayreuth}\\
\baselineskip=14pt
{\small\em D-95440 Bayreuth, Germany}\\}\baselineskip=16pt~\\
{\bf Ola T\"{o}rnkvist\footnote{Electronic address:
{\tt O.Tornkvist@damtp.cam.ac.uk}}}\\ {
{\small\em NASA/Fermilab Astrophysics Center}\\
\baselineskip=14pt
{\small\em MS-209, P.O.~Box 500, Batavia, IL
60510-0500, U.S.A.}\\
{\small\em and}\\
{\small\em Department of Applied Mathematics and Theoretical Physics}\\
{\small\em University of
Cambridge, Cambridge CB3 9EW,
United Kingdom\footnote{Present address.}}\\}
\baselineskip=16pt
\vspace*{.75cm}
{19 November 1997}
\end{center}
\vspace{0.2cm}
\begin{abstract}
We derive an exact equation of motion for a non-relativistic vortex
in two- and three-dimensional models with a complex field. The velocity
is given in terms of gradients of the complex field at the vortex
position. We discuss the problem of reducing the field dynamics to
a closed dynamical system with non-locally interacting
strings as the fundamental degrees of freedom.
\end{abstract}
\newpage
\renewcommand{\thefootnote}{\arabic{footnote}}
\setcounter{footnote}{0}
\baselineskip=16pt
\section{Introduction}
Vortices are extended solitonic objects that are (locally)
stable solutions in
many classical field theory models
involving a complex field $\Phi=
|\Phi| e^{iS}$.
In three dimensions they resemble strings, but have a
core of finite width where the fields deviate appreciably from
their asymptotic values.
The complex field $\Phi$ is zero
on a one-dimensional string located inside the core.
Along this string the phase $S$ is multivalued, and
singlevaluedness of $\Phi$ implies that the
line integral $\oint\!dS$
along a contour enclosing the string is $2\pi n$
where $n$ is the integer winding number of the vortex.
Stable vortices with
$n>1$ will generally have $n$ distinct strings of
zeros located within the core. One can therefore
regard such a multivortex as a bound state of unit-,
or elementary, vortices with $n=1$. For the special case
of a single, stationary,
straight, infinitely long,
isolated, and therefore cylindrically symmetric
vortex solution, it is well-known that $\Phi\sim r^{|n|} e^{in\varphi}
\equiv z^n$ $({z^{\ast}}^{(-n)})$, where $r$ is the
perpendicular distance to the string.

Non-relativistic models
with vortex solutions include the Ginzburg-Pitaevski-Gross
model for superfluid $^4$He, with global U(1) symmetry, and the
time-dependent Ginzburg-Landau model of a superconductor, with
local U(1) symmetry. These are non-dissipative
models whose field equations can be
derived from Lagrangians. Their relativistic generalizations
are known as the Goldstone model and abelian Higgs model, respectively,
whose vortex solutions constitute global and gauged cosmic strings.

Non-relativistic vortices are solutions to equations with
first-order time derivatives, which are typically of the form
\begin{equation}
\label{geneq}
%\dot
\frac{d\Phi}{dt}= b\,\nabla^2 \Phi + P(\Phi,\Phi^{\ast})\Phi~,
\end{equation}
where $b\in {\bf C}$ and $P$ is a polynomial. When $ib\notin
{\bf R}$, the system is dissipative and no Lagrangian description
exists. Vortex solutions in such a
system have the
distinctive feature that the lines of constant phase $S$
are spirals rather than radial rays. Correspondingly, in three dimensions
the sheets of constant phase are rolled up as a scroll
around the central string. Such spiral vortex
solutions occur in a variety of physical systems, such as
chemical reaction-diffusion experiments, thermal
convection, the growth of slime mold, non-linear laser optics, and
even in the human heart.
\vspace*{-2mm}
\section{String Formulation}

The idea behind this work is to treat the string of zeros of $\Phi$
as a fundamental object, and to  obtain an exact equation of motion for
the string in terms of the fields that surround it. This is possible
because the string is a feature of a {\em local\/} field theory,
and therefore its motion can be determined from the behaviour of
fields in an infinitesimal neighbourhood of the string. All the
necessary information is encapsuled in the underlying field theory.

Because the strings have no thickness, the equation of motion for
the string is exactly valid
for an arbitrarily small intervortex distance
as well as for arbitrarily large (but finite) vortex curvature.
This is of considerable interest in the study of situations
when vortex cores overlap, such as vortex-vortex scattering,
bound multivortex states, cusp formation, the intersection and
reconnection of vortex segments, collapsing string loops, and
the question of vortex rigidity.
In the string formulation, the strings interact non-locally through
fields that are not confined to worldsheets but
live in four-dimensional space-time.
\vspace*{-2mm}
\section{The String Equation of Motion}

The principal steps in deriving the string equation of motion
are most clearly demonstrated in the case
of  non-relativistic vortices \cite{us}
that satisfy a field equation of the form
(\ref{geneq}).
The equation for relativistic vortices has been worked
out by Ben-Ya'acov \cite{BenY}.

Inserting $\Phi\equiv Re^{iS}$ into eq.~(\ref{geneq})
with $b = b_{\scriptscriptstyle R} + i b_{\scriptscriptstyle I}$,
 one obtains
the amplitude and phase equations
\begin{equation}
\label{Aabseq}
{{d}\over{dt}} \ln R= {\rm Re}(P) +
b_{\scriptscriptstyle R}
\left(\frac{1}{R}\nabla^2 R - (\nabla S)^2\right)
-{{b_{\scriptscriptstyle I}}\over{R^2}}
\nabla\cdot (R^2 \nabla S)~~~
\end{equation}
\begin{equation}
\label{Seq}
{{d}\over{dt}} S = {\rm Im}(P) + b_{\scriptscriptstyle I}
\left(\frac{1}{R}\nabla^2 R
- (\nabla S)^2\right)
+{{b_{\scriptscriptstyle R}}
\over{R^2}} \nabla\cdot (R^2 \nabla S)~.~~
\end{equation}
Local curvilinear coordinates are then introduced
in a neighbourhood of
the string by writing $\bm{x} = \bm{X}(s,t) +
x \bm{N}(s,t) + y \bm{B}(s,t)$ where $s$ is the arc-length
along the string,
$\bm{X}(s,t)$ is the position of the string at a given time $t$,
and \bm{N},\,\bm{B} are the normal and binormal unit vectors,
respectively. An arbitrary position $\bm{x}$ near the string
is thus
specified at any time $t$ by the coordinates $s,x,$ and $y$. Local
polar coordinates are defined by $x=r\cos\varphi$,
$y=r\sin\varphi$.

One can now work out the gradient and Laplacian operators in
the curvilinear coordinates. They depend on the curvature
$\kappa$ and torsion $\tau$ of the string. The time
derivative in eqs. (\ref{Aabseq}) and (\ref{Seq}) is
expressed as
\begin{equation}
\frac{d\,}{dt} = - (\dot{\bm{X}} +
 x \dot{\bm{N}} + y \dot{\bm{B}})
\cdot \nabla + \frac{\partial\,}{\partial t}~,
\end{equation}
where $\dot{\bm{X}} = \partial{\bm{X}}/\partial t$
etc., and $\partial/\partial t$ indicates the time derivative
in a frame following the local segment of the string such that
$s,x,y$ are constant.

The equation of motion for the string is derived
by identifying the parts of  eqs.~(\ref{Aabseq}) and (\ref{Seq})
that are singular as $r\to 0$ and demanding that these
singularities cancel order by order. To this end, let us
separate out the non-differentiable
parts of $R$ and $S$, writing $\ln R = \ln R_{\rm sing.} +
\ln w$ and $S = S_{\rm sing.} + \theta$, where $\ln w$ and $\theta$
are everywhere differentiable. The calculations are simplified
if we choose  a ``gauge''
where the singular parts are given by $R_{\rm sing.}= r^{|n|}$
and $S_{\rm sing.}=n\varphi$ as for the straight, isolated
vortex solution. The final result is independent
of this restriction \cite{us}.

The singular
terms of order $r^{-2}$ are easily shown to cancel in both
equations. Terms of order $r^{-1}$ on the left-hand side of the
equations contain the string velocity $\dot{\bm{X}}$, while
those on the right-hand side do not. In this way
an equation for $\dot{\bm{X}}$
is obtained. For example,
$dS/dt$ contains the term $ -\dot{\bm{X}}\cdot \nabla S_{\rm sing.}
= -(n/r)\,\hat{\bm{\varphi}}\cdot \dot{\bm{X}}$, and
$(\nabla S)^2$ contains the term $\nabla S_{\rm sing.}\cdot
\nabla\theta = (n/r)\,\hat{\bm{\varphi}}\cdot \nabla\theta$.
Collecting all terms of order $r^{-1}$
in both equations one obtains the following expression for
the string velocity:
\begin{eqnarray}
\label{veq}
{\dot{\mbox{\bm{X}}}}&=&
{b_{\scriptscriptstyle I} \left(\kappa {{n}\over{|n|}}
\mbox{{\bm{B}}} +
2\nabla_\perp{\theta}-
2{{n}\over{|n|}}\mbox{{\bm{T}}}\times
\nabla\ln {w} \right)}\nonumber\\*
&+&{b_{\scriptscriptstyle R}\left(\kappa \mbox{\bm{N}} -
2\nabla_\perp\ln w -
2\frac{n}{|n|}\mbox{{\bm{T}}}\times\nabla\theta
\right)}{\hspace{.15cm} {\rm ,}\hspace{-.15cm}}
\end{eqnarray}
where $\bm{T}=\partial\bm{X}/\partial s$ is the
string tangent vector and
\mbox{$\nabla_\perp = -\bm{T}\times (\bm{T}\times \nabla)$} is the
gradient projection perpendicular to the string.
Gradients are to be evaluated at the position of the string.
The two-dimensional result is obtained as $\kappa\to 0$.

The velocity can also be expressed
in terms of the original magnitude $R$ and phase $S$.
Because of an exact cancellation of singularities,
this expression is identical to
eq.~(\ref{veq}) with the substitutions
$\theta\to S$, $\ln w \to
\ln R$. A beautifully compact expression is obtained by defining
$\dot{Z}\equiv (\bm{N} + i\bm{B})\cdot \dot{\bm{X}}$ and
$z=x+iy$. Then one finds
$\dot{Z} = b\,[-4\,\partial(\ln \Phi)/\partial z^{\ast} + \kappa]$
or $\dot{Z}^{\ast} = b \,[-4\,\partial(\ln \Phi)/\partial z + \kappa]$
for positive and negative $n$ respectively, where the right-hand
sides are evaluated on the string $(z=0)$. From the
physical requirement of finite string velocity one then
infers that the field $\Phi$ near the string is either
holomorphic or anti-holomorphic.

The result shows that, apart from Biot-Savart
terms proportional to curvature,
the string velocity expression includes local
gradients of the magnitude and phase of the complex field $\Phi$.
A similar result is obtained for the non-dissipative relativistic
vortex,
satisfying the equation \cite{BenY}
\begin{equation}
\label{releq}
X^\mu{}^{;a}_{;a} = -\frac{n}{|n|} \epsilon^\mu{}_{\nu\lambda\rho}
X^\lambda_{,a} X^\rho_{,b}\epsilon^{ab}\sqrt{-\gamma}\,\partial^\nu\theta
+ 2 (g^{\mu\nu} - \gamma^{ab} X^\mu_{,a} X^\nu_{,b})\,
\partial_\nu \ln w~,
\end{equation}
where $\sigma^a$  ($a=0,1$) are coordinates on the
worldsheet and $\gamma^{ab}$ is the induced worldsheet metric.
Eq.~(\ref{releq}) reduces to the Nambu-Goto equation for a free
relativistic string when the right-hand side is zero.
\vspace*{-2mm}
\section{Conclusions and Outlook}

We have shown that the motion of a vortex string,
defined as the set of zeros of the complex field $\Phi$,
can be expressed exactly in terms
of gradients of the magnitude and phase of $\Phi$ at the position of
the string. This represents only partial progress towards a
formulation of vortex dynamics in terms of strings, since $\Phi$
satisfies a
non-linear partial differential equation whose solutions are not known.
Further developments require approximations. Near
the Bogomolnyi limit of 2D gauge theories
one may make an adiabatic approximation in which slowly moving vortices
at each instant take on the field configuration of known static solutions
\cite{Nick}. The field dynamics can then be reduced to a
finite-dimensional system of
differential equations for the vortex positions.
Other approaches involve perturbations of the straight, isolated
vortex solution, the small parameters being e.g.\ the curvature,
$b_{\scriptscriptstyle R}/b_{\scriptscriptstyle I}$, the core radius
\cite{others}. Such approximations usually include only local
contributions to the dynamics
or make specific assumptions about the asymptotic field behavior.
 We emphasise that the equation derived here is
exact and incorporates
also long-range contributions from other vortices and remote segments
of the same vortex.

\subsection*{Acknowledgments}
One of us (O.T.) thanks the conference organisers for their splendid
work.
Support for O.T. was provided in part by the Swedish
Natural Science Research Council (NFR)
and by DOE and NASA under Grant NAG5-2788, in part
by EPSRC under Grant GR/K50641. E.S. was supported by the Danish
Natural Science Research Council. 
\bibliographystyle{unsrt}

\begin{thebibliography}{1}
\setlength{\parsep}{0pt}
\setlength{\parskip}{0pt}
\bibitem{us}
O.~T\"{o}rnkvist and E.~Schr\"{o}der,
Phys.\ Rev.\ Lett.\  {\bf 78} (1997) 1908.

\bibitem{BenY}
U.~Ben-Ya'acov,
 Nucl.\ Phys.\ {\bf B382} (1992) 597.

\bibitem{Nick}
N.~Manton, Ann.\ Phys.\ {\bf 256} (1997) 114.

\bibitem{others}
M.~Gabbay, E.~Ott and P.N.~Guzdar,
Phys.\ Rev.\ Lett.\ {\bf 78} (1997) 2012;
I.S.~Aranson and A.R.~Bishop,
{\em Stretching of vortex lines and
generation of vorticity in the three-dimensional complex
Ginzburg-Landau
Equation\/},
preprint patt-sol/9705009.
\end{thebibliography}

\end{document}